# Chemical induced delithiation on Li$_x$MnPO$_4$: an investigation about the phase structure

Melanie Köntje

ZSW (Zentrum für Sonnenenergie- und Wasserstoff-Forschung, Baden-Württemberg),
Helmholtzstrasse 8, 89081 Ulm, Germany

Giorgia Greco[1]

Helmholtz-Zentrum Berlin für Materialien und Energie GmbH, Hahn-Meitner-Platz 1, D-14109 Berlin, Germany.
giorgia.greco@helmholtz-berlin.de

Giuliana Aquilanti

ELETTRA-Sincrotrone Trieste,
34149 Basovizza (TS), Italy
giuliana.aquilanti@elettra.eu

Luca Olivi

ELETTRA-Sincrotrone Trieste,
34149 Basovizza (TS), Italy
luca.olivi@elettra.ue

Peter Axmann

ZSW (Zentrum für Sonnenenergie- und Wasserstoff-Forschung, Baden-Württemberg),
Helmholtzstrasse 8, 89081 Ulm, Germany
peter.axmann@zsw-bw.de

Margret Wohlfahrt-Mehrens

ZSW (Zentrum für Sonnenenergie- und Wasserstoff-Forschung, Baden-Württemberg),
margret.wohlfahrt-mehrens@zsw-bw.de

[1] Correspondig author

# ABSTRACT

Understanding the LiMnPO$_4$/MnPO$_4$ phase transition is of great interest in order to further improve the electrochemical performance of this cathode material. Since most of the previously published literature deals with characterization of chemically delithiated Li$_x$ MnPO$_4$, the aim of this study is to compare and study the composition and structure of the different phases that are gen-erated upon chemical delithiation of Li$_x$ MnPO$_4$. Bare and carbon-coated lithium manganese phos-phates are prepared via a combined coprecipitation-calcination method. Partial delithiation to two different degrees of delithiation Li$_x$ MnPO$_4$ (x = 0.24/0.23 and 0.45) for carbon-coated and/or bare materials is achieved using an excess of nitro-nium tetrafluoroborate in acetonitrile. The ef-fect of carbon-coating has been also considered. Standard materials characterization with XRD (X-Ray Diffraction) and ICP-OES (Inductive Coupled Plasma spectrometry and Optical Emission Spectroscopy) analysis are in accordance with literature data, but further cerimetric analysis revealed seri-ous deviations, showing differences in the degree of delithiation to the average degree of oxidation. A structural characterization of the atomic and electronic local structure of the materials is also ob-tained using XAS (X-ray Absorption Spectroscopy) technique.

# Keywords

Energy storage, High Voltage Cathodes, Phase Transition, XRD, XAFS, Phospho Olivines, Batteries, Chemical Delithiation.

# 1. INTRODUCTION

Transition metal phospho olivines turned out to be a promising alternative to the commonly used layered oxides since Padhi et al. first reported about the electrochemical performance of LiFePO$_4$ as possible cathode material [16]. In comparison to the commonly used commercialized lay-ered oxides phospho olivines provide a low toxicity- environmental benignity, higher potentials (3.45 V vs. Li/Li$^+$ for LiFePO$_4$, 4.1 V vs. Li/Li$^+$ for LiMnPO$_4$) [21, 23] and thermal stability [24]. Many efforts have been made to investigate the electrochemical de-/ intercalation of lithium ions into the LiFePO$_4$ host pronouncing a first-order heteroge-neous two phase transition [16, 20, 18]. For the promising lithium manganese phospho olivine the phase transition has not been investigated in such detail, but the LiMnPO$_4$/ MnPO$_4$ transition is announced to proceed in an equivalent heterogeneous two phase transition as investigated for the iron analogue [2, 5]. The lower electrochemical performance in comparison to the iron analogue is commonly addressed to the inactivated Mn$^{2+}$/Mn$^{3+}$ redox reaction. This phenomenon has been mostly explained by a lower electronic and ionic conductivity, interface strain at the LiMnPO$_4$/ MnPO$_4$ phase boundary and local distortions due to the Jahn-Teller-ion Mn$^{3+}$ [23, 15, 4, 17]. Reporting detailed infor-mation about the phase composition, the Li$_x$ (Mn$_y$Fe$_{1-y}$)PO$_4$ (0≤x≤1) phase diagram first published by Yamada et al. [25] shows distinct regions of solid solutions and two phase regimes. Further detailed structural investigations of the partly delithiated Li$_x$ MnPO$_4$ materials pointed out effects due to differences in the electronic states of Mn$^{3+}$-O$_6$ and Fe$^{3+}$-O$_6$ which cause the Jahn-Teller distortion for Mn$^{3+}$ In the present paper we aim to evaluate and compare the composition of the different phases that generate upon chemical delithiation of the Li$_x$ MnPO$_4$.

For the purpose to characterize these products we used a number of spectroscopic and analytical techniques. XRD and ICP-OES analysis turned out to well agree with previ-ously published literature. Additionally we took advantages of cerimetric analysis, a powerful technique that helped to gain further details on the oxidation state of Mn. We show severe differences for the average degree of oxidation mea-sured by cerimetry to the calculated value from ICP-OES analysis. Furthermore a XAS analysis has been performed, allowing us a full structural characterization including infor-mation about oxygen bonding and valence state.

# 2. EXPERIMENTHAL

## 2.1 Sample Preparation

Synthesis of the carbon-coated and non carbon-coated lithium manganese phospho olivines were performed via a combined coprecipitation calcination route. The precursor was obtained from appropriate aqueous solutions containing the metal salts in certain molar ratios. Annealing the precur-sor in an inert atmosphere resulted in the desired LiMnPO$_4$ phospho olivines. For the carbon coating appropriate organic carbon containing solid substances were introduced to the precursor and properly mixed before annealing. The amount therefore has been calculated to obtain 5wt% carbon relative to the amount of manganese on the LMPO/C. Chemical delithiation of the bare and uncoated LiMnPO$_4$ has been achieved by applying excess nitroniumtetrafluoroborate in acetonitrile (4:1 relative to the amount of Mn(II), dry reaction conditions, inert gas, elevated temperature 40ºC) under kinetic reaction control. To stop the delithiation process the LMPO/C material has been separated from the delithiation agens, accurately washed and dried. Reaction duration of 29min (Li0 45MnPO4/C), 90min (Li0 24MnPO4/C).

and 21min (Li0 24MnPO4) have been applied Mn(II) was determined by cerimetry and thus the conversion Mn(II) to Mn(III) calculated. A Methrom Titrino titration system has been used. A more detailed description is reported elsewhere [1]. We have obtained the materials listed below.

Carbon coated samples:

LiMnPO$_4$ labeled as LMPO/C;

Li$_{0.45}$MnPO$_4$ labeled as LMPO-0.5/C;

Li$_{0.24}$MnPO$_4$ labeled as LMPO-0.2/C;

Uncoated sample:

Li$_{0.23}$MnPO$_4$ labeled as LMPO-0.2.

## 3. RESULTS AND DISCUSSION

### 3.1 Chemical delithiation procedure of LMPO

Since pure Li$_x$MnPO$_4$ powders could be easily obtained via several chemical delithiation procedures [25, 13] these materials have been often investigated for detailed examina-tions of the LiMPO$_4$/MPO$_4$ phase transition [24, 2, 25, 13, 12].
The delithiation process as described in the synthesis procedure exhibits significant differences compared to elec-trochemical delithiation procedures. During electrochemi-cal delithiation the redox process takes place at electrically conducted points at the active materials and additives grains within the electrolyte. The reaction is mainly dependent from charge-transfer characteristics, ion diffusivity and the SEI characteristics of this electrode composition. Contrarily, the chemical delithiation process takes not place within the standard electrolyte environment considering the SEI, but takes place in non-aqueous solution at the complete grain surface. The reaction is expected to directly take place be-tween the active materials grains and the delithiation agens with the lack of SEI characteristics and in changed elec-trolyte to common standard electrochemical conditions. The obtained chemically delithiated phases are supposed to be not identical to electrochemically delithiated Li$_x$MnPO$_4$ as we will show during the material characterization.

### 3.2 Compositional analysis: degree of oxidation of samples

Firstly, the degree of oxidation of manganese in the pure LMPO/C has been evaluated, thus having a benchmark. The expected calculated value from ICP-OES analysis of 2.00 has been confirmed by cerimetry (2.01), as shown in Fig. 1 (x = 1). Contrarily both chemically delithiated carbon-coated Li$_x$MnPO$_4$ (x = 0.45 and x = 0.24) showed lower values in comparison to the calculated degrees of oxidation of manganese. This hence indicates a formal higher amount of "Mn$^{2+}$" concomitant to a lower amount of "Mn$^{3+}$" as already considered. Due to these differences we suppose that the assumption all residual lithium depicts the equivalent amount of Mn$^{2+}$ which crystallises in the olivine structure, is not complete. Thus it should be extended by an amount of amorphous Mn$^{2+}$ containing phosphates with no retaining lithium in the structure. Delithiated Mn$^{III}$PO$_4$ is known to decompose to manganese(II)pyrophosphate under oxygen release during thermal treatment [2, 12, 3]. The formal Li:Mn:P ratio does not underlie any changes caused by the MnPO$_4$ to Mn$_2$P$_2$O$_7$ transformation. Subsequently there could not be detected any change in the chemical composition via ICP-OES analysis. During the chemical delithiation process we suppose an equivalent decomposition of the MnPO$_4$ phase. Since we found a degradation of the material caused by the chemical delithiation procedure we suppose the formation of amorphous Mn$^{2+}$ containing phases not detectable for XRD as reason for the deviations depicted from ICP and cerimetry.

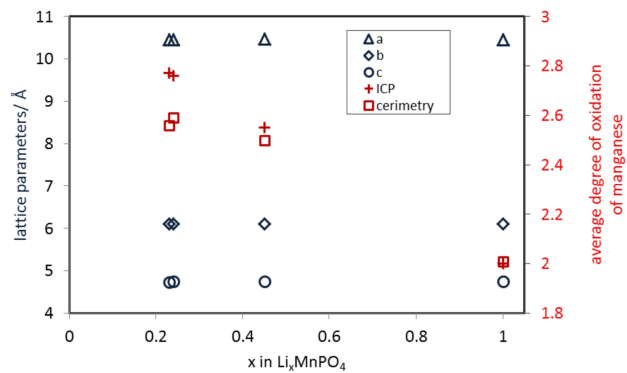

Figure 1: Right scale: comparison of the average degree of oxidation of manganese calculated via ICP-OES analysis and cerimetry for LMPO/C, LMPO-0.2 and LMPO-x/C (x = 0.5 and 0.2). Left scale: the x lattice paramenters *a*, *b* and *c* (spacegroup *Pnma*) in Å obtained by XRD analysis.

### 3.3 Structural characterization with XRD technique

The chemical composition was calculated from ICP-OES analysis and normalized to the amount of manganese in LMPO-0.5/C and LMPO-0.2/C. Powder diffraction pat-terns of LiMnPO$_4$ carbon-coating starting material LMPO/C and of the chemically delithiated ones are shown in Fig. 2 as well as the non carbon-coated chemically delithiated one (magenta curve).
The X-ray diffraction pattern for the LMPO/C (red curve) shows the peaks of the orthorhombic LiMnPO$_4$ olivine (*Pnma*). From the quantitative Rietveld analysis, we could determine also a rather low amount of Li$_3$PO$_4$ (2.3 wt%). Further quantification of the two phases, LiMnPO$_4$ and Li$_3$PO$_4$, has been done by ICP-OES chemical analysis, which revealed 97.6 wt% LiMnPO$_4$ and 2.4 wt% Li$_3$PO$_4$ in full accordance with the XRD results. As it can be seen from the X-ray patterns of the chemically delithiated sam-ples (Fig. 2, blue and red curves), an additional phase of MnPO$_4$ with purpurite structure appears under consumption of LiMnPO$_4$. No Li$_3$PO$_4$ could be indexed, which we ex-pect to be partially extracted during the chemical delithia-tion process. Nevertheless phase quantities less than 2 wt% are below the limit of determination via XRD. The delithi-ation process leads to a broadening of the peaks related to the Li$_x$MnPO$_4$ phases. We therefore evaluated the lattice parameters of the remaining lithiated phase (Fig.1) and cal-culated the crystallite sizes of the carbon-coated Li$_x$MnPO$_4$ according to Pawley-fits. All lattice parameters remain un-changed, showing negligibly higher values with increasing degree of oxidation. Tab 1 reports the comparison of the rel-evant values obtained. Both chemically delithiated LMPO-x/C (x = 0.5 and x = 0.2) exhibit a reduction of the crystallite size for the remaining lithiated phase to ~1/2, and for the new delithiated phase of ~1/3 compared to the raw mate-rial. The results reported until now are in well agreement with previously published literature data [24, 2, 25, 1, 12], where chemically delithiated $Li_x MnPO_4$ have been obtained via treatment with nitronium tetrafluoroborate.

The influence of the carbon coating has been also evaluated. The carbon coating has been added with the intention of creating an additional protecting and supporting surface layer on the active material considering the electronic conductivity during the redox process as well as the lithium ion diffusion mechanism during the in-/exsertion process. In Fig 2 (upper panel, magenta curve) the diffraction pattern of LMPO-0.2 is shown. It is obvious that the carbon-coating prevents the material of a great amorphization. This might be due to the strongly enhanced reaction rate of delithiation for the non carbon-coated material, as already pointed out in section 2.1.

The XRD results reported in Fig. 2 for the chemi-cally delithiated samples demonstrated the coexistence of a fully lithiated phase ($LiMnPO_4$) and of a fully delithiated phase ($MnPO_4$). This would be explained by a heteroge-neous two phase transition. Since the $Li_xMnPO_4$ stoichiom-etry has been determined on the basis of ICP-OES analy-sis, and considering that no impurities can be detected from the X-ray patterns, we believe that all the residual lithium-ions reside in the olivine structure. Thus the stoichiometry of Li0 45MnPO4 leads to a phase composition of 45 mol%.

$LiMnPO_4$ and 55 mol% $MnPO_4$ which refers to an average degree of conversion of 0.55. Similarly, the $Li_{0.24}MnPO_4$ material refers to an average degree of conversion of 0.76. From these significant results the average degree of oxidation of manganese has been calculated. The two degrees of conversion of 0.55 and 0.76 lead to the average degrees of oxidation of manganese of 2.55 and 2.76 within the heterogeneous $LiMn^{II} PO_4$/ $Mn^{III} PO_4$ system.

Results from cerimetry show for the non carbon-coated LMPO-0.2 in comparison to the carbon-coated equivalent, even a greater deviation to the calculated value of the average degree of oxidation (Fig. 1). Considering its higher degree of amorphization concomitant to this deviation, we assume for the non carbon-coated material even a higher decomposition of the chemically delithiated $MnPO_4$.

Table 1: Crystallite sizes in terms of diameters d obtained by XRD analysis for pristine LMPO/C and chemically delithiated carbon coated samples.

| | LMPO/C Raw | LMPO-0.5/C LiMnPO4 | MnPO4 | LMPO-0.2/C LiMnPO4 | MnPO4 |
|---|---|---|---|---|---|
| Size d (nm) | 94 ± 5 | 51 ± 10 | 34 ± 10 | 53 ± 7 | 29 ± 6 |

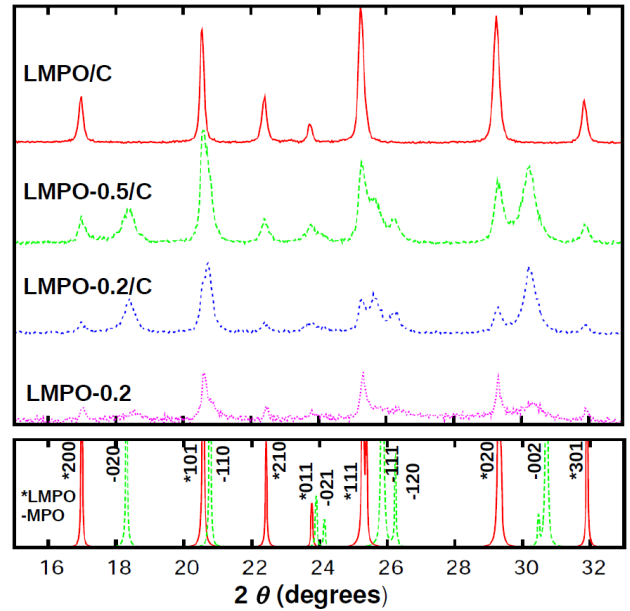

Figure 2: Upper panel: powder diffraction patterns of chemically delithiated LMPO-x/C with carbon-coating (x = 0.5, 0.2) and without LMPO-0.2. Lower panel: theoretical diffraction patterns of the lithiated phase of LiMnPO4 (labeled as LMPO) and completed delithiated phase MnPO4 (labeled as MPO).

### 3.4  Local structure investigation with XAS technique

.As it is well known, the XAS technique is sensitive to the local atomic structure up to 5-10 Å around photoabsorbing sites, selected by their atomic number, and therefore it is a very useful tool for nanostructural and amorphous ma-terials study [9]. Due to the possibility to extract chemi-cally specific short range information, in combination with XRD it allows a full structural characterization including in-formation about bonding and valence state. Regarding the investigations shown above of the average degree of oxidation of manganese in the following samples LMPO-0.5/C and LMPO-0.2 are investigated more in detail for the fol-lowing reason. LMPO-0.5/C and LMPO-0.2/C show similar values for the crystallite sizes. The duration of their delithiation processes have been in relation to each other. The deviation of the average degree of oxidation of manganese within the ICP based values and cerimetric results differ for the LMPO-0.2/C more, LMPO-0.5/C are closer. Contrarily, the uncoated LMPO-0.2 suffers from quite high parti-cle size degradation and an even higher deviation between the ICP-OES based values and cerimetry results. From this point of view we aim to focus on LMPO-0.5/C considering its smallest deviations within the evaluation techniques, and the LMPO-0.2 which shows the greatest deviations and material degradation. Combining them enables to give insight into the effect of the carbon coating first and second the manganese content e.g. its phases composition after a chemical delithiation process.

### 3.4.1 X-ray Absorption Near Edge Spectroscopy (XANES)

In Fig. 3(a), we report the near-edge XAS spectra of the Mn K-edges of LMPO-0.5/C and LMPO-0.2 after removal of the pre-edge absorption (without normalization to the absorption jump). The amount of Mn quantity is different for the two samples. After the delithiation procedure the amount of Mn in the LMPO-0.2 sample is lower of ~ 60% respect to the LMPO-0.5/C, but the electronic structure is mainly the same for the different delithiated samples as shown in Figure 3(b) by the direct comparison between the normalized near-edge XAS spectra. This is also confirmed by the XANES of the Mn K-edge pre-edge position proper of the second order 1s–3d transition shown in Fig. 4. The pre-edge and the Mn K-edge positions are first qualitative indicators of the valence of manganese [19, 11]. For both LMPO-0.5/C and LMPO-0.2 a similar valence state between II and III can be confirmed. Moreover, looking at Fig. 4 a slight shift of the absorption maximum towards higher energy and an increase of intensity are indices for a slight increase of the Mn valence state for the more delithiated sample LMPO-0.2. Comparing this result with the above described compositional analysis, it can be seen that the mean increased degree of oxidation of manganese for the LMPO-0.2 sample evaluated by ICP-OES analysis and cerimetry is confirmed by XAS analysis. It should be noted that there is no mean difference in the Mn K-edge position for the two samples.

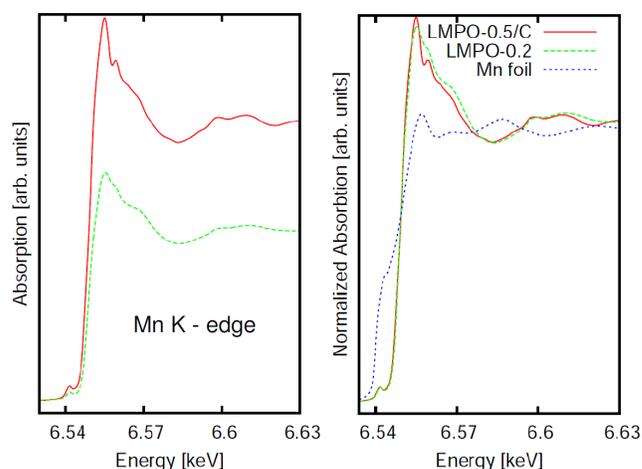

**Figure 3:** (a) Comparison of XANES Mn K-edge spectra of LMPO/C with different delithiation state powder, (b) the XANES normalized spectra of the samples are compared with the Mn foil at K - edge

### 3.4.2 X-ray Absorption Near Edge Spectroscopy (XANES)

The LMPO-0.5/C and LMPO-0.2 extracted EXAFS signals and the moduli of their Fourier Transforms (normalized) are compared in Fig. 5 and first shell fitting is shown in Fig. 6, while first shell detailed results of GNXAS structural refinements are presented in Tab 2. Looking at EXAFS signals at Fig. 5 upper panel, the difference in the intensity is evident, due to the mainly amorphous nature of the LMPO-0.2 sample. Looking at the normalized moduli of the Fourier Transforms (Fig. 5 , bottom panel) the LMPO-0.2 intensity of furthers shells signals appear less intense indicating in this sample only a short range order. Three different Mn-O distances were found for LMPO-0.5/C and LMPO-0.2 (Fig. 6) in agreement with crystallographic and vibrational data [6] and lattice parameters obtained by XRD. The delithiated phase $Mn^{3+}PO_4$ (here mentioned as MPO) could be fitted with two different Mn-O distances of coordination number 4 (square planar) and 2 (vertical arrangement). This clearly shows the non-uniform $Mn^{3+}$-$O_6$ environment due to the Jahn-Teller distortion. For the residual fully lithiated $LiMnPO_4$ (LMPO) phase EXAFS analysis shows only one Mn-O distance indicating an uniform Mn-O coordination in the $Mn^{2+}$-$O_6$ coordination polyhedral. The more delithiated LMPO-0.2 sample shows an equivalent EXAFS analysis with three Mn-O distances correlated to the $Mn^{3+}$-$O_6$ and $Mn^{2+}$-$O_6$ polyhedra. But a Mn-O (I) contraction and increasing structural disorder (Debye-Waller parameter) could be obtained (see Fig. 6 and Tab. 2).

The deviations to smaller values in comparison to literature data still needs to be further worked on. Also the different behavior of the Mn-O (I) and Mn-O (II) distances should be further investigated. This could be further discussed in light of the assumed formation of amorphous $Mn^{2+}$ containing phases.

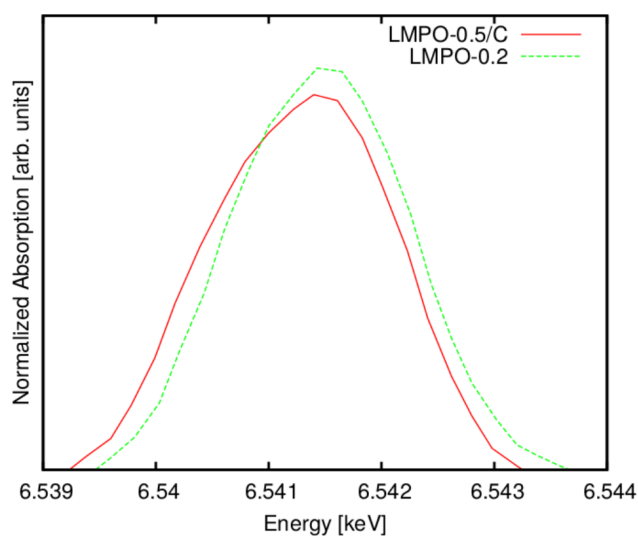

**Figure 4:** Normalized Mn pre-edge feature of the two samples with different delithiation state.

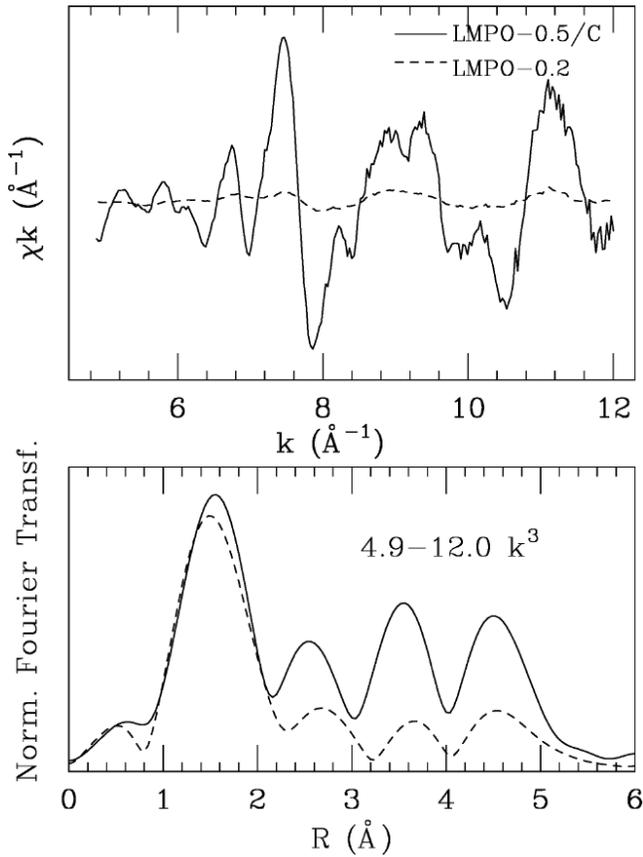

**Figure 5:** Mn K-edge EXAFS spectra (after background subtraction, *k* weighted) (top) and Fourier Transforms (FT) (bottom) for LMPO-0.5 solid line, LMPO-0.2 dashed line.

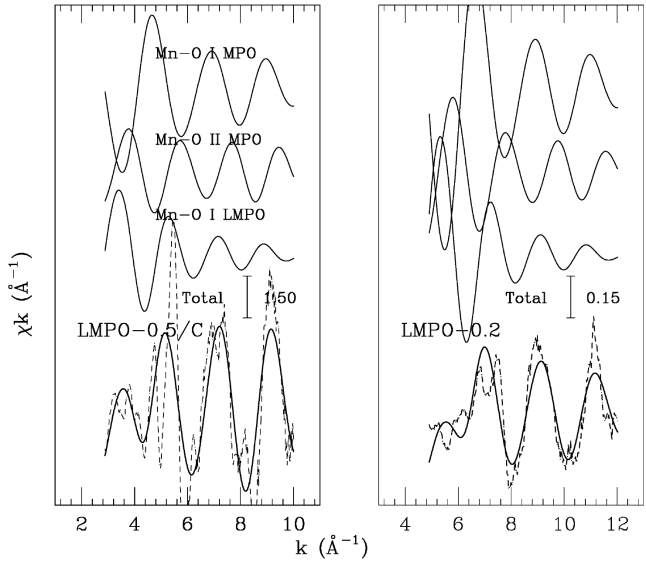

**Figure 6:** The best fit results of GNXAS analysis performed for LMPO-0.5/C and LMPO-0.2 Mn K-edge ($k^3$ weighted).

Upper curves represent components of the model signal:

Mn-O total two-body signal for the LMPO/C and MPO phases.

## 4. CONCLUSIONS

The above shown investigations aim to clarify the conformity of chemically and electrochemically delithiated $Li_x MnPO_4$. Experimental values from cerimetric analysis gave reasons to out doubts about the direct comparability of chemically and electrochemically delithiated $Li_x MnPO_4$. The lower average degree of oxidation examined by cerimetry leads to the assumption of decomposition of the delithiated phase $MnPO_4$ and formation of amorphous $Mn_2P_2O_7$ or generally spoken $Mn^{2+}$-containing amorphous phosphorous rich phases during the chemical delithiation process since no other crystalline phases are apparent in XRD analysis. High-quality XAS spectra at the Mn K-edge of the LMPO-0.5/C and LMPO-0.2 were show the presence of amorphous $Mn_2P_2O_7$ can be by the XANES pre-peak shape and posi-tion and by a deeper EXAFS analysis evaluating the signal contribution proper of the $Mn_2P_2O_7$ phase. From XANES and EXAFS analysis confirmed valence states between II and III, in agreement with to cerimetry. Moreover LMPO-0.2 sample shows a lower content of Mn respect to the less delithiated LMPO-0.5/C. During the delithiation process the 60% of Mn is lost due to the solubility of $Mn^{2+}$ A weak short range order indicating the strong material grinding and amorphization for LMPO-0.2 was shown. From first shell fitting first detailed results of GNXAS structural refinements have been presented. For both investigated samples a strong Jahn-Teller distortion, proper of the delithieted phase, has been detected. Deviations of the bond lengths from literature still need further investigations facing the expected decomposition of $MnPO_4$. Furthermore, facing XRD and ICP-OES analysis additional characterization techniques like cerimetry would be proposed. The obtained results will be used in our future works.

**Table 2:** Structural parameters obtained by GNXAS analysis: $R_{fit}$ [Å] - the average inter-atomic distance, $\sigma^2$ [$10^{-3}$ Å$^2$] - the standard deviation of distance, the $N_x$ (x=cal and tot) - is the obtained and total coordination number respectively. $R_{model}$ [Å] is the expected interatomic distance

|  | bond I-shell | $R_{model}$ | $R_{fit}$ | $\sigma^2$ | $N_{cal}$ |
|---|---|---|---|---|---|
| **LMPO-0.5/C:** | | | | | |
| (1-x) MnPO$_4$ | Mn-O I | 1.95 | 1.88(5) | 8.0(1) | 4* |
| (1-x) MnPO$_4$ | Mn-O II | 2.08 | 2.07(3) | 3.0(1) | 2* |
| x LiMnPO$_4$ | Mn-O I | 2.22 | 2.17(5) | 10(5) | 6 |
| **LMPO-0.2:** | | | | | |
| (1-x) MnPO$_4$ | Mn-O I | 1.95 | 1.85(2) | 5.7(5) | 4* |
| (1-x) MnPO$_4$ | Mn-O II | 2.08 | 2.05(3) | 3.0(1) | 2* |
| x LiMnPO$_4$ | Mn-O I | 2.22 | 2.14(5) | 11(1) | 6 |

*$N_{tot}$ (2+4)=6 (Jahn-Teller dist.)

## 5. ACKNOWLEDGMENTS

This work is financially supported by the Deutsche Forschungsgemeinschaft DFG (WO882/4-2). The authors gratefully thank the XAFS beamline service group at Synchrotron Light Laboratory Elettra (XAFS Station, Trieste, Italy) for their support. We wish also to thank Mrs. G.